\documentclass[aps,prl,letterpaper,amsmath,amssymb,superscriptaddress,nofootinbib,nolongbibliography,twocolumn]{revtex4-2}

\usepackage{graphicx}
\usepackage[usenames]{color}
\usepackage[colorlinks=true,linkcolor=magenta,citecolor=magenta,anchorcolor=green,urlcolor=magenta]{hyperref}
\usepackage{mathrsfs}

\usepackage[percent]{overpic}
\usepackage{tabularx}
\usepackage{multirow}
\usepackage{orcidlink}

\usepackage{CJK}
\newcommand\w           {\omega}
\newcommand{\wec}[1]   {|#1\rangle}
\newcommand{\cew}[1]   {\langle#1|}

\def\be{\begin{equation}}
\def\ee{\end{equation}}
\def\ba{\begin{aligned}}
\def\ea{\end{aligned}}
\def\bea{\begin{eqnarray}}
\def\eea{\end{eqnarray}}

\begin{document}
\begin{CJK*}{UTF8}{}
\title{Magnetization oscillations in a periodically driven transverse field Ising chain}
\author{Xiao Wang
\CJKfamily{gbsn}(王骁)~\orcidlink{0000-0003-2898-3355}}
\affiliation{Tsung-Dao Lee Institute,
Shanghai Jiao Tong University, Shanghai, 201210, China}
\affiliation{Shanghai Branch, Hefei National Laboratory, Shanghai 201315, China}

\author{Masaki Oshikawa \CJKfamily{gbsn}(押川正毅)~\orcidlink{0000-0002-7637-7432}}
\affiliation{Institute for Solid State Physics, The University of Tokyo. Kashiwa, Chiba 277-8581, Japan}
\affiliation{Kavli Institute for the Physics and Mathematics of the Universe (WPI), The University of Tokyo, Kashiwa, Chiba 277-8583, Japan}
\author{M\'{a}rton Kormos~\orcidlink{0000-0002-1884-2818}}
\altaffiliation{kormos.marton@ttk.bme.hu}
\affiliation{Department of Theoretical Physics, Institute of Physics, Budapest University of Technology and Economics, 1111 Budapest, M{\H u}egyetem rkp. 3, Hungary}
\affiliation{HUN-REN--BME Quantum Dynamics and Correlations Research Group, Budapest University of Technology and Economics, 1111 Budapest, M{\H u}egyetem rkp. 3, Hungary}
\author{Jianda Wu \CJKfamily{gbsn}(吴建达)~\orcidlink{0000-0002-3571-3348}}
\altaffiliation{wujd@sjtu.edu.cn}
\affiliation{Tsung-Dao Lee Institute,
Shanghai Jiao Tong University, Shanghai, 201210, China}
\affiliation{Shanghai Branch, Hefei National Laboratory, Shanghai 201315, China}
\affiliation{School of Physics \& Astronomy, Shanghai Jiao Tong University, Shanghai, 200240, China}

\begin{abstract}
We investigate the nonequilibrium dynamics of the magnetization in an Ising chain subjected to a slowly rotating transverse field.
The magnetization oscillations are found to be explained by the contributions from different particle
excitations
in the quantum $E_8$ model.
We study the magnetization in the frequency domain in detail,
uncovering a series of singular peaks for the $z$ (Ising) component.
These singular peaks are split into two sets for the magnetization along
$x$ and $y$ directions
with frequency shifts
set by the
rotational-field frequency.
The peaks include both $\delta$-function type and edge-singularity type peaks.
The $\delta$-function peaks can be attributed to particle excitations involving an $E_8$ particle with either the
vacuum or a different
particle.
The edge-singularity peaks are contributed by particle excitations of two $E_8$ particles with either the
vacuum or another
particle, or by particle excitations that contain two sets of two
particles
with each set including at least
a particle of the same type.
We propose a Rydberg qubit array for possible experimental investigation.
\end{abstract}

\maketitle
\end{CJK*}

\textit{Introduction.}---Time-dependent quantum systems have attracted continuous attention in the past few decades, leading to rich experimental and theoretical progress.
These include time crystals~\cite{PhysRevLett.109.160401,PhysRevLett.109.160402,PhysRevLett.114.251603,caizi2021}, nonlinear electron spin resonance and photon-induced phase transitions~\cite{PhysRevLett.82.5136,PhysRevB.65.134410,iwai,okamoto,kitagawa,oka2009},
measurement induced exotic phenomena~\cite{PhysRevX.9.031009,PhysRevResearch.2.013022,PhysRevB.108.165120,PRXQuantum.4.030317,masaki2023}, as well as Kibble-Zurek mechanism~\cite{Quan_2010,kh2020,PhysRevB.110.045140,PhysRevA.109.053320,PhysRevB.109.064501}.
Along the course
the transverse-field quantum Ising chain (TFIC) serves as a prototypical
system for investigating out of equilibrium dynamics \cite{CalabreseEsslerFagotti,Calabrese_2012I,Calabrese_2012II}.
For a static transverse field, the system is one of the simplest solvable models that accommodates quantum phase transition~\cite{Pfeuty,sachdev_2011}.
At its quantum critical point (QCP), the system exhibits conformal invariance~\cite{Daniel}.
Introducing a perturbation field along the Ising-spin direction at the QCP
gives rise, in the scaling limit, to an
integrable field theory known as quantum $E_8$ integrable model~\cite{zam,DELFINO1995724, zhao,zou,xiao,jiahao_review,xiao_review,xiao_2023}.
When the transverse field is time dependent,
a complete analytical understanding poses a significant challenge.

There are two different widely-studied forms of the time-dependent magnetic field.
One is a sudden change of the magnetic field at the initial time~\cite{PhysRevLett.110.257203,CalabreseEsslerFagotti,marton2016,Delfino_2017,kh_2018,Etienne2020}, and the other is a periodically driven magnetic field~\cite{Robinson,oka2009,oka2014}.
For a sudden quench involving a nonzero longitudinal field, the dynamics of both the one-point functions and the entanglement entropy reveal fingerprints of the confinement phenomenon~\cite{marton2016,Delfino_2017,kh_2018,H_ds_gi_2019,Alessio2020,Knolle2021,Stefan2022,Anna2024}.
For the periodic driving, it has been proved that for
a rotating field in the transverse plane, the time evolution can be related to the TFIC
with the presence of an effective longitudinal field proportional to the rotation frequency~\cite{oka2014,Robinson}.

In this letter, we focus on the critical Ising chain with a periodically driven magnetic field that slowly rotates in the transverse plane.
Under a rotational wave transformation~\cite{rwa_07},
the time evolution process can be effectively related to a time-independent critical TFIC
perturbed by an effective longitudinal field proportional to the rotation frequency~\cite{Robinson}.
In the scaling limit, the effective model becomes the quantum $E_8$ integrable model~\cite{xiao}.
Using the exact form factors  of the model and the post-quench perturbative framework~\cite{DELFINO1995724,marton2010, H_ds_gi_2019,kh2020,xiao, olallaEE, olallaEEjhep, kh_2018}, we analytically determine
the time evolution of the magnetization components $M^\alpha$ $(\alpha = x, y, z)$ in both
the time and frequency domains, where the dominant
low-energy
spectrum can be understood through contributions from different
particle states of the quantum $E_8$ integrable model.

For the magnetization spectra $M^{z}(\omega)$, we find that
the low-energy collective excitations in frequency domain are
dominated by cascades of singular peaks of two different types:
the $\delta$-function type and the edge-singularity type.
The former singular peaks are contributed by the
particle excitations of an $E_8$ particle
with either the  vacuum or a different particle. The latter
are due to the particle excitations of two  particles with either the  vacuum
or an $E_8$ particle, or due to the particle excitations of two sets of
two particles, where
the two sets contain particles of the same type
~\cite{SM}.
For the magnetization spectra $M^{x}(\omega)$ and $M^{y}(\omega)$,
cascades of singular peaks also appear with similar structure and the
same origin, but split into two sets
with respect to the peaks of $M^{z}(\omega)$,
with their corresponding frequency
shifts matching the rotating frequency of the transverse field.
Numerical verification using the infinite time-evolving block decimation (iTEBD) algorithm is also presented, supporting the analytical results.
Finally, a Rydberg-atom quantum simulator is proposed for future experimental study,
providing a promising way to simulate and investigate
the interplay of non-equilibrium physics and quantum integrability.

\textit{Model.}---The Hamiltonian of a critical Ising chain subjected to a
periodically driven transverse field $\vec{B}(t) = ({\cos(\w_{0} t)}$, $-\sin(\w_{0} t), 0)$
is given by
\begin{equation}
    \mathcal{H}=-J\sum_{i}\left[\sigma^{z}_{i}\sigma^{z}_{i+1}+\cos(\w_{0} t)\sigma^{x}_{i}-\sin(\w_{0} t)\sigma^{y}_{i}\right],
\label{Eq:LatH}
\end{equation}
where $\sigma_{i}^{\alpha}(\alpha=x,~y,~z)$ represent
the Pauli matrices at site $i$, and we assume that the rotation
frequency $(\hbar)\omega_{0}\ll J$.
Hereafter, $\hbar$ and $J$ are set as 1. At $t=0$, the system is a critical TFIC,
and we set the initial state to be its ground state.
By using a unitary transformation~\cite{oka2014,Robinson,Quan_2010,SM},
\begin{equation}
\mathcal{U}_{D}(t)=\exp\left\lbrace -\frac{i}{2}\w_{0} t\sum_{i}\sigma_{i}^{z}\right\rbrace,
\label{Eq:latUnitary}
\end{equation}
the time-evolved wave function $\wec{\psi(t)}$ is given by
\begin{equation}
    \wec{\psi(t)}=\mathcal{U}_{D}^{-1}(t)e^{-i\mathcal{H}_{\text{eff}}t}\wec{\psi(0)}
    \label{Eq:quenchlatevo}
\end{equation}
with the effective static Hamiltonian
\begin{equation}
    \mathcal{H}_{\text{eff}} =-\sum_{i}\left(\sigma_{i}^{z}\sigma_{i+1}^{z}+\sigma_{i}^{x}-\frac{1}{2}\w_{0}\sigma_{i}^{z}\right),
\label{Eq:quenchH}
\end{equation}
which is a quantum critical TFIC perturbed by an effective longitudinal field $\w_{0}/2$.
We consider the time-dependent magnetization along the $x$, $y$, and $z$ directions,
\begin{equation}
    M^{\alpha}(t)=\langle \sigma^{\alpha}(t)\rangle =\lim_{N\rightarrow\infty} \frac{1}{N}\sum_{i=1}^{N}\langle \sigma_{i}^{\alpha}(t)\rangle.
    \label{Eq:spincompoents}
\end{equation}
Introducing the time-dependent magnetic components $D^{\alpha}(t)$ as
\begin{equation}
    D^{\alpha}(t)=\langle\psi(0)|e^{iH_{\text{eff}}t} \sigma^{\alpha} e^{-iH_{\text{eff}}t}|\psi(0)\rangle,
    \label{Eq:Mcomponents}
\end{equation}
the time-dependent magnetization $M^{\alpha}(t)$ components are
\begin{equation}
    \begin{aligned}
        M^{x}(t)&=\cos(\w_{0}t)D^{x}(t)+\sin(\w_{0}t)D^{y}(t),\\
        M^{y}(t)&=\cos(\w_{0}t)D^{y}(t)-\sin(\w_{0}t)D^{x}(t),\\
        M^{z}(t)&=D^{z}(t).
    \end{aligned}
    \label{Eq:Mtimedomain}
\end{equation}
It is in general difficult to determine Eq.~(\ref{Eq:Mcomponents}) analytically.
However, for small $\w_{0}$ the effective Hamiltonian Eq.~(\ref{Eq:quenchH})
turns out to be the quantum $E_{8}$ Hamiltonian in the scaling limit~\cite{zam,DELFINO1995724,xiao},
\begin{equation}
    \mathcal{H}_{E_{8}} = \mathcal{H}_{c=1/2}+h\int dx\sigma(x).
\label{Eq:HE8}
\end{equation}
Here, $\mathcal{H}_{c=1/2}$ represents the Hamiltonian of conformal field theory (CFT) with central charge $c=1/2$.
The coupling intensity $h$ corresponds to $\w_{0}/2$ in the scaling limit.
The Pauli matrices $\sigma^{z}$ and $\sigma^{x}$ correspond to
the primary fields of the Ising and energy density fields $\sigma$ and $\varepsilon$, respectively.
In the quantum $E_8$ integrable model, the matrix elements
of these two primary fields can be determined using the form factor bootstrap method~\cite{DELFINO1995724,H_ds_gi_2019,xiao,SM}. Taking the scaling limit of $D^{\alpha}(t)$ [Eq.~(\ref{Eq:Mcomponents})]
\be
\begin{aligned}
\label{Eq:Ds}
\mathscr{D}^{\mathcal{O}}(t) = \sum_{n,m=0}^{\infty}A^{\mathcal{O}}_{nm}(t)
\end{aligned}
\ee
for $\mathcal{O}=\varepsilon,\sigma$, hereafter the math script font of corresponding physical quantity refers to the scaling limit. We have inserted two complete $E_8$ bases $\{\vert n;j\rangle \}$ and
$\{\vert m;k\rangle \}$ where we organize the states according to their particle numbers $n$ and $m$ and $j,k$ label states within these sectors. Then $A^{\mathcal{O}}_{nm}(t) = \sum_{j,k}e^{i(E_{n;j}-E_{m;k})t}\langle\psi_{s}(0)\vert n;j\rangle \langle n;j\vert\mathcal{O}\vert m;k\rangle \langle m;k\vert\psi_{s}(0)\rangle$, with $\vert\psi_{s}(0)\rangle$ being $\vert\psi(0)\rangle$ in the scaling limit.
In addition, $E_{n;j}$ and $E_{m;k}$ are eigenenergies of the two $E_8$ eigenstates $\vert n;j\rangle$ and $\vert m;k\rangle$, respectively.
The overlaps $\langle m;k \vert\psi_{s}(0)\rangle$ can be determined using a post-quench perturbative framework~\cite{kh_2018,H_ds_gi_2019,kh2020,SM},
and the explicit form of the $A^{\mathcal{O}}_{nm}(t)$ can be found in~\cite{SM}. In the following, we call the terms in Eq.\ \eqref{Eq:Ds} with fixed particle numbers particle channels.
The remaining challenge is to determine $D^{y}(t)$ in the scaling limit, as $\sigma^{y}$ does not correspond to a primary field in the Ising CFT. However, starting from the lattice formalism, we notice~\cite{jianda_2014,xiao,SM}
\begin{equation}
    D^{y}(t) = \frac{1}{2}\frac{d D^{z}(t)}{dt}\stackrel{\text{scaling limit}}{\longrightarrow}\mathscr{D}^{y}(t) = \frac{1}{2}\frac{d \mathscr{D}^{\sigma}(t)}{dt}.
    \label{Eq:calcuYs}
\end{equation}
Therefore, by taking the system back to the scaling limit
we can simultaneously determine $\mathscr{D}^{\sigma}$ and $\mathscr{D}^{y}$.

\begin{figure*}[tp]
    \includegraphics[width=17.5cm]{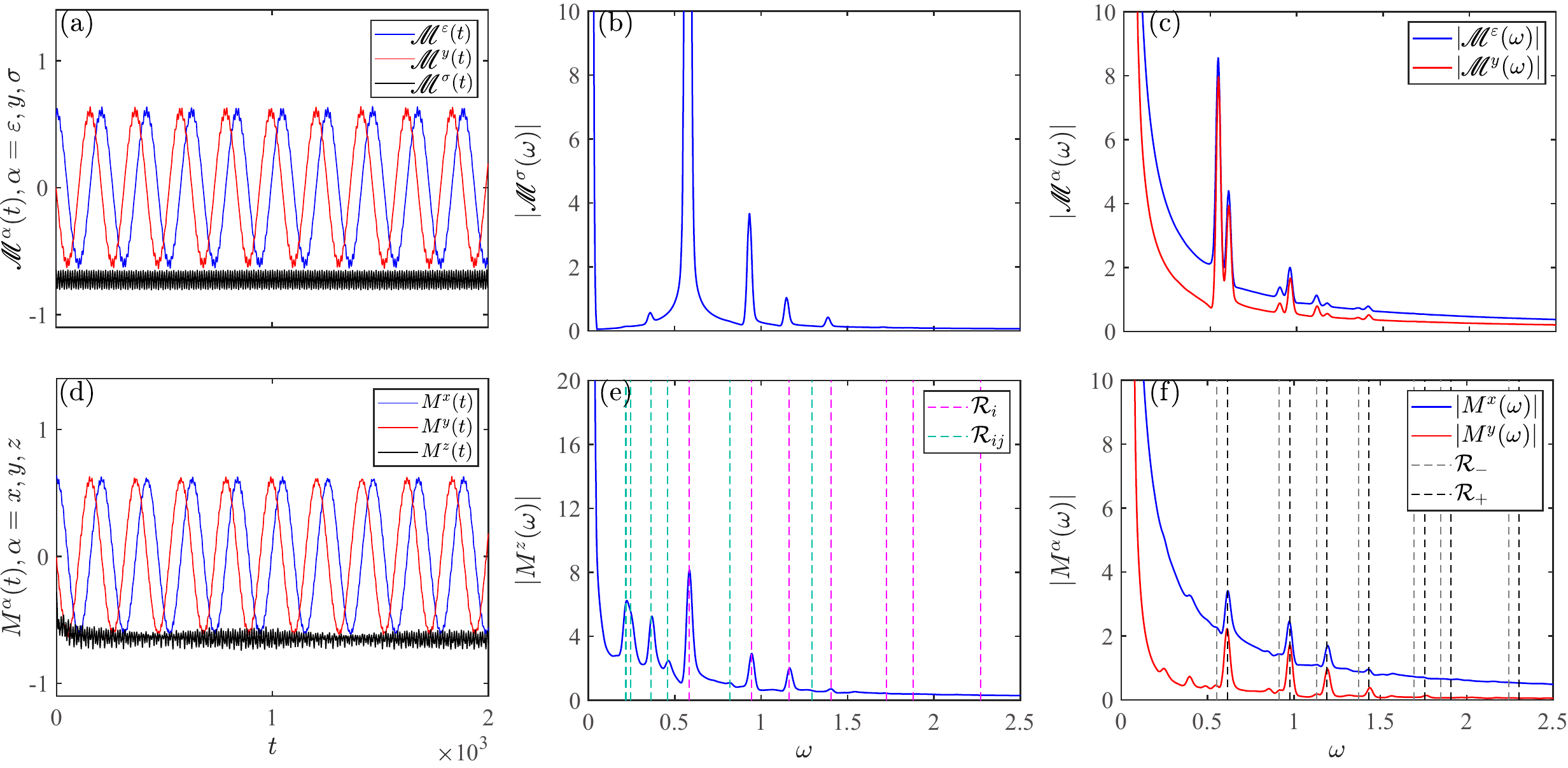}
    \caption{ (a), (d) Time-dependent magnetization calculated in the quantum $E_8$ field theory [$\mathscr{M}^{\alpha}(t),\alpha=\varepsilon,y,\sigma$] and iTEBD simulation for the lattice model [Eq.~(\ref{Eq:LatH}), ($M^{\alpha}(t),\alpha=x,y,z$)], respectively. (b), (e) $|\mathscr{M}^{\sigma}(\omega)|$ and $|M^{z}(\omega)|$, respectively. For better illustration, $\mathscr{M}^{\sigma}(t)-\langle\sigma\rangle$ is enlarged with a factor 10. Correspondingly the spectral weights for $\w>0$ in $|\mathscr{M}^{\sigma}(\w)|$ are also enlarged with this factor. (c), (f) $|\mathscr{M}^{\varepsilon,y}(\omega)|$ and $|M^{x,y}(\omega)|$, respectively. In (e), the dashed magenta and green lines label singular peaks coming from channels of an $E_8$ particle with either the $E_8$ vacuum ($\mathcal{R}_{i}$) or a different $E_8$ particle ($\mathcal{R}_{ij}$), with detailed frequencies listed in Table.~\ref{table:mzfrequency}. In (f) the two groups of singular peaks $\mathcal{R}^{+}$ (black dashed line) and $\mathcal{R}^{-}$ (grey dashed line) for single $E_8$ particles with corresponding frequency $m_{\mathcal{R}^{\pm}}=m_{\mathcal{R}}\pm\omega_0$. NOTE: For the numerical results the peak positions match the analytical results accurately. Due to the finite truncated dimension for long-time evolution in iTEBD algorithm, the spectral weights at low-frequency ($\omega < 1$) should not be compared with analytical ones. For the analytical results due to heavy cost of multi-particle form factor calculations, some multi-particle's contribution are not shown in the figure.}
    \label{fig:spectra}
\end{figure*}

\textit{Singular peaks.}---Using Eq.~(\ref{Eq:Mtimedomain}) and Eq.~(\ref{Eq:calcuYs}), the results for $\mathscr{M}^{\mathcal{O}}(t)$, with $\mathcal{O}=\varepsilon, y$, and $\sigma$, are displayed in Fig.~\ref{fig:spectra} (a), after normalization with $m_1=C_{\w}{\w_{0}}^{8/15}$, $\langle\sigma\rangle=-C_{\sigma}{\w_{0}}^{1/15}$ and $\langle\varepsilon\rangle=\langle\varepsilon\rangle_{h=0}-C_{\varepsilon}{\w_{0}}^{8/15}$. Here we set $\w_0=0.03$, and the explicit values of $C_{\w}$, $C_{\sigma}$ and $C_{\varepsilon}$ can be found in \cite{SM}. To better understand the collective excitations in the time-dependent magnetization, we consider
magnetization in the frequency domain following the one-sided Fourier transformation,
\begin{equation}
    \mathscr{M}^{\mathcal{O}}(\w)=\int_{0}^{\infty} dt e^{i\w t} \mathscr{M}^{\mathcal{O}}(t).
    \label{Eq:onesideFFMs}
\end{equation}

We present our analytical results alongside those obtained from
numerical iTEBD simulations in Figure~\ref{fig:spectra}.
To facilitate a clearer comparison between the analytical and numerical outcomes,
we applied Gaussian broadening with a broadening factor of
$\alpha=2\%$.
The detailed analytical expressions are provided in \cite{SM}.
As can be seen in the figure, the dominant singular peaks predicted by the analytical field-theoretical approach
are verified by the numerical simulations,
which, overall, are given by contributions from single- and two-$E_8$-particle channels in the frequency domain~\cite{H_ds_gi_2019,SM}.
For $\mathscr{M}^{\sigma}(\w)$ [$M^{z}(\w)$ for the lattice simulation], cascades of singular peaks $\{\mathcal{R}\}$ are found in the low-energy excitations.
Such singular peaks are corresponding to two different types of singular behaviors: the $\delta$-function type and the edge-singularity type.
For the $\delta$-function peaks, besides those peaks from an $E_{8}$ particle and the $E_8$ vacuum, other peaks emerge from the contributions of two different $E_8$ particles, with the corresponding frequency being their mass difference~\cite{marton2016,kh_2018,olallaEE,SM}.
For the edge-singularity peaks, they have two distinct origins.
The first one is due to the divergence in the density of states of $E_8$ particles, also known as Van Hove singularity~\cite{xiao,mussardo_E7}.
Such edge-singularity peaks are contributed by two different $E_8$ particles $m_{i},~m_{j}$, and the other one being either the $E_8$ vacuum or a different $E_8$ particle $m_{k}$, showing a singular behavior as $1/\sqrt{\w-\w_\text{th}}$, where $\w_\text{th}=m_{i}+m_{j}$ (or $=m_{i}+m_{j}-m_{k}$) being the threshold~\cite{SM}.
The second one originates from the kinematic singularities of the
multi-particle form factor of the quantum $E_8$ field theory~\cite{SM,DELFINO1995724,H_ds_gi_2019,xiao,mussardo_E7},
associated with the particle channels of two distinct sets of
$E_8$ particles.
One set comprises two $E_8$ particles $m_i$ and $m_k$, while the other set
includes
two $E_8$ particles with at least one of the same type as $m_j$ and $m_k$.
These contribute $\ln(\w+m_{i}-m_{j})$ edge-singularities
in addition to the previous $\delta$-function peaks
at the thresholds of $\w_\text{th} =m_{j}-m_{i}$.
It is worth noticing that
in the particle channels of a single
to two $E_8$ particles, the kinematic poles also appear
and seem to contribute additional singularities,
which will eventually be
cancelled due to the exchange property of the multi-particle form factors
~\cite{SM}.
In the numerical simulation of the Hamiltonian Eq.~(\ref{Eq:LatH}),
by comparing the corresponding frequencies of the
peaks with the analytical predictions,
the dominant contributions are identified as
$\delta$-function peaks~[dashed vertical lines in
Fig.~\ref{fig:spectra} (e)],
with their detailed frequencies listed in Table.~\ref{table:mzfrequency}.
Additionally, other singularity peaks are
identified as edge-singularity peaks contributed
by the vacuum-to-two-particle channels of $m_{1}m_{2}$,
the one-to-two-particle channels of $m_{5}-m_{1}m_{4}$ and $m_{1}m_{2}-m_{5}$~\cite{SM}.
However, for the later two classes,
the heavy $E_8$-particle channels are not shown in the analytical results
due to the high computational cost of form factor calculations.

\begin{figure*}[t]
    \includegraphics[width=17.5cm]{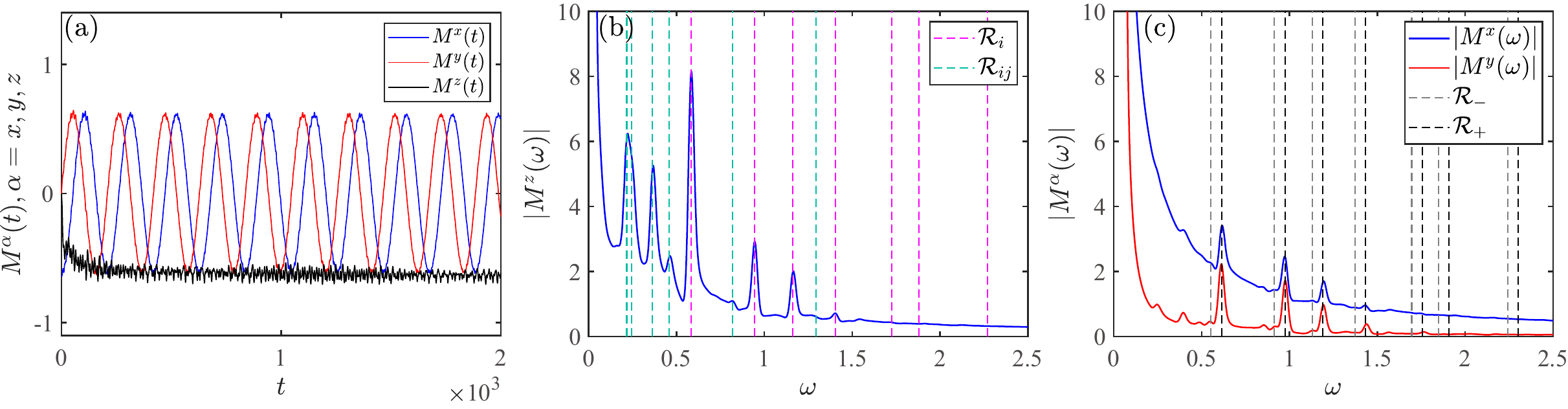}
    \caption{ (a) iTEBD simulation on the magnetization of the Rydberg Ising Hamiltonian [Eq.~(\ref{Eq:longIsing}), ($M^{\alpha}(t),\alpha=x,y,z$)]. (b, c) $|M^{z}(\omega)|$ and $|M^{x,y}(\omega)|$. All the peaks have the
     same structure as Fig.~\ref{fig:spectra}(e,f).}
    \label{fig:coldatom}
\end{figure*}

\begin{table*}
\caption{Ratios of the magnetization peaks in frequency domain correspond to single $E_8$ particles (1$^{\text{st}}$ row) or the combinations of two $E_8$ particles (2$^{\text{nd}}$ row) are given in the 3$^{\text{rd}}$ row along with their expected values for a specific rotation frequency $\w_0=0.03$ (4$^{\text{th}}$ row). The peaks simulated by iTEBD algorithm for $\w_0=0.03$ according to the original lattice model [Eq.~(\ref{Eq:LatH})] and the proposed Rydberg Ising Hamiltonian [Eq.~(\ref{Eq:longIsing})] are also listed (5$^{\text{th}}$ and 6$^{\text{th}}$ row), shown in Fig.~(\ref{fig:spectra}) (d) and Fig.~(\ref{fig:coldatom}) (b), as the vertical dashed lines in the spectra, respectively.}
\centering
\begin{tabular}{p{0.12\linewidth}p{0.06\linewidth}p{0.06\linewidth}p{0.06\linewidth}p{0.06\linewidth}p{0.05\linewidth}p{0.06\linewidth}p{0.05\linewidth}p{0.05\linewidth}p{0.06\linewidth}p{0.05\linewidth}p{0.06\linewidth}p{0.05\linewidth}p{0.05\linewidth}}
\hline \hline
One particle & & & & & $m_1$ & & $m_2$ & $m_3$ & & $m_4$ & & $m_5$ & $m_6$ \\
\hline
Two particles &$m_3$-$m_2$&$m_4$-$m_3$&$m_2$-$m_1$&$m_4$-$m_2$& & $m_4$-$m_1$ & & &$m_6$-$m_1$& &$m_1$+$m_2$  \\
\hline \hline
Theoretical ratio $m_i$/$m_1$ & 0.3710 & 0.4158 & 0.6180 & 0.7868 & 1 & 1.4049 & 1.6180 & 1.9890 & 2.2183 & 2.4049 & 2.6180 & 2.9563 & 3.2183 \\
\hline
Theoretical frequency & 0.2139 & 0.2398 & 0.3563  & 0.4537 & 0.5766 & 0.8100 & 0.9329 & 1.1468 & 1.2790 & 1.3866 & 1.5095  & 1.7045 & 1.8556\\
\hline
lattice simulation & 0.2238 & 0.2238 & 0.3597 & 0.4563 & 0.5763 & 0.8024 & 0.9293 & 1.1485 & 1.2700 & 1.3809 & 1.5166 & 1.6989 & 1.8628\\
\hline
Rydberg simulation & 0.2166 & 0.2427 & 0.3608 & 0.4593 & 0.5838 & 0.8201 & 0.9446 & 1.1611 & 1.2950 & 1.4039 & 1.5283 & 1.7258 & 1.8788\\
\hline
\end{tabular}
\label{table:mzfrequency}
\end{table*}

In the spectra of $\mathscr{M}^{\varepsilon}(\w)$ and $\mathscr{M}^{y}(\w)$,
the stable singular peaks $\{\mathcal{R}\}$ are further separated into two distinct groups of new singular peaks $\{\mathcal{R}^{\pm}\}$, where the corresponding excitation frequencies have been shifted as $m_{\mathcal{R}}\pm\w_0$, shown in Fig.~\ref{fig:spectra} (c). This special feature is also captured by the numerical
simulation for $M^{x}(\w)$ and $M^{y}(\w)$ [Fig.~\ref{fig:spectra} (f)], where the separated single $E_8$ particles are also highlighted by dashed vertical lines. The numerical verification agrees well with the analytical results.

\textit{Implementation in a Rydberg qubit array}.---
For possible experimental study of the $E_8$ physics in a periodically driven system, we propose a many-body time-dependent Rydberg Hamiltonian~\cite{Aquila, coldatom1},
\begin{equation}
\begin{aligned}
    \mathcal{H}_{Ry}(t)&=\frac{\Omega(t)}{2}\sum_{i}e^{i\phi(t)}\wec{g_{i}}\cew{r_{i}}+e^{-i\phi(t)}\wec{r_{i}}\cew{g_{i}} \\
    &-\Delta(t)\sum_{i}n_{i}+\sum_{i<j}\frac{C_{6}}{|x_{i}-x_{j}|^{6}}n_{i}n_{j},
\end{aligned}
\label{Eq:Aquila}
\end{equation}
where $\wec{r_{i}}=\wec{1_{i}}$ and $\wec{g_{i}}=\wec{0_{i}}$ are the
representation of the Rydberg states and the ground states on the
$i^{th}$ qubit, respectively. The amplitude and phase of the driving laser field is $\Omega(t)$ and $\phi(t)$, respectively.
The operator $n_{i}=\wec{r_{i}}\cew{r_{i}}$ counts the Rydberg excitations, and $x_{i}=ia$ are
the coordinate for the $i$-th qubit with $a$ being the lattice spacing.
$\Omega(t)$, $\phi(t)$, $\Delta(t)$ and $C_{6}$ are the corresponding (time-dependent) coefficients,
whose values can be found in Ref. \cite{Aquila}. By applying a transformation between the Rydberg states and an effective Ising basis
$\sigma_{i}^{+}=\wec{g_{i}}\cew{r_{i}},~\sigma_{i}^{-}=\wec{r_{i}}\cew{g_{i}},~\sigma_{i}^{z}=\mathbf{1}-2n_{i}$, this Rydberg Hamiltonian can  be transformed to a periodically driven transverse field Ising chain.
Since the interaction strength decays as $1/|i-j|^6$, we neglect spin interaction term with distance larger than 2 lattice sites, so the effective Hamiltonian follows~\cite{coldatom1},
\begin{equation}
\begin{aligned}
    &\mathcal{H}_{\text{Ising}}/{|C_{6}|}=\\
    &-\sum_{i}\sigma^{z}_{i}\sigma^{z}_{i+1}+\lambda\sigma^{z}_{i}\sigma^{z}_{i+2}-g_{c}[\cos(\w_{0}t)\sigma^{x}_{i}-\sin(\w_{0}t)\sigma^{y}_{i}],
\end{aligned}
\label{Eq:longIsing}
\end{equation}
where $\lambda=1/2^6$.
Using iTEBD algorithm, we find that the model hosts a quantum phase transition
at the quantum critical point $g_{c}=1.0275$,
which also falls into the TFIC universality class.
The rotating frequency is set as $\w_0=0.03$. All the corresponding numerical results are shown in Fig.~\ref{fig:coldatom} (a), (b) and (c),
with a Gaussian broadening $\alpha=2\%$.
The dominated low-energy physics proposed by the quantum $E_8$ integrable field theory are well
captured by the numerical lattice simulation results, and the corresponding peaks
are also shown in Table.~\ref{table:mzfrequency}. It can be found that the additional
next nearest neighboring Ising interaction
only slightly modifies the frequency of the singular peaks.
Our numerical simulations on Eq.~(\ref{Eq:longIsing})
implies that the $E_8$ physics revealed in the time-dependent model [Eq.~(\ref{Eq:LatH})]
is fully accessible through a practical Rydberg qubit array experimental setup.

\textit{Conclusions}.---In this letter, we showed that the time
evolution of the magnetization in the Ising chain under a slowly rotating transverse field can be
described via the integrable $E_8$ quantum field theory and its particle content.
Based on a post-quench perturbation framework,
along with the $E_{8}$ form factor theory,
we analytically determined the magnetizations in both the time and frequency
domains.
Along the $z$ direction, we identified two types of singular peaks in the frequency domain: the $\delta$-function and the edge-singularity types.
The frequencies of the $\delta$-function peaks
are given by the masses and mass differences of $E_{8}$ particles.
The edge-singularity peaks are essentially the Van Hove singularities
given by the divergence of the density of states, or the kinematic singularities of the multi-particle $E_8$ form factors. A significant contribution given by the particle combination
$m_{1}m_{2}$ is found in the numerical data.
Moreover, for the
magnetizations along the $x$ and $y$ directions,
in the frequency domain all the singular peaks are
split into two sets of new peaks with their corresponding frequency
shifts matching the rotation frequency.

For a possible experimental implementation, we
proposed a neutral atom Rydberg quantum simulator that can realize a modified periodically
driven transverse field Ising chain. We confirmed numerically
that despite the possible presence of next-nearest-neighbor couplings, it can realize the expected magnetization spectra,
thus providing a novel and practical route to access
the quantum $E_8$ physics in a time-dependent system.

Switching on a static transverse field provides an exciting opportunity to study the dynamics of confinement, which we leave for future work. Our work may also inspire further studies of non-equilibrium physics via quantum integrability.

\textit{Acknowledgements.}---We thank Ning Xi, Zongping Gong and Zilong Li for discussions, Congjun Wu for helpful discussion during an early stage of the work.
The work at Shanghai Jiao Tong University is sponsored by National Natural Science Foundation of China No. 12274288 and the Innovation Program for Quantum Science and Technology Grant No. 2021ZD0301900. MK was supported by the National Research, Development and Innovation
Office of Hungary (NKFIH) through the OTKA Grant K138606. The work of M.O. was partially supported by the JSPS KAKENHI Grants No. JP23K25791 and No. JP24H00946.

\bibliography{main}

\newpage
\appendix
\onecolumngrid

\section{\large{Supplemental Material---Magnetization oscillations in a periodically driven transverse field Ising chain}}

\section{A.~The emergent quantum $E_{8}$ integrable model}
\label{app:A}
In this appendix, we review the derivation of the effective static Hamiltonian and its scaling limit originally given in Ref. \cite{Robinson}. We start from the original periodically driven transverse field Ising chain to derive the emergent quantum $E_{8}$ model. The Schr$\ddot{\rm{o}}$dinger equation is
\begin{equation}
    i\partial_{t}\vert\psi(t)\rangle =\mathcal{H}(t)\vert\psi(t)\rangle.
    \label{S1}
\end{equation}
To solve this equation, considering a time-dependent unitary transformation,
\begin{equation}
i\partial_{t}\mathcal{U}_{D}(t)\vert\psi(t)\rangle =\left[ \mathcal{U}_{D}(t)\mathcal{H}(t)\mathcal{U}_{D}^{-1}(t)+i\dot{\mathcal{U}_{D}}(t)\mathcal{U}_{D}^{-1}(t)\right]\mathcal{U}_{D}(t)\vert\psi(t)\rangle, 
\label{S2}
\end{equation}
with
\begin{equation}
\mathcal{U}_{D}(t)=\exp\left\lbrace -\frac{i}{2}\w_0 t\sum_{i}\sigma_{i}^{z}\right\rbrace.
\label{S3}
\end{equation}
Substituting Eq.~(\ref{S3}) into Eq.~(\ref{S2}), obtaining
\begin{equation}
\mathcal{U}_{D}(t)\mathcal{H}(t)\mathcal{U}_{D}^{-1}(t)+i\dot{\mathcal{U}_{D}}(t)\mathcal{U}_{D}^{-1}(t)=-\sum_{i}(\sigma_{i}^{z}\sigma_{i+1}^{z}+\sigma_{i}^{x})+\frac{1}{2}\w_0 \sum_{i}\sigma_{i}^{z}.
\label{S4}
\end{equation}
In the scaling limit with $\w_0\sim 0$, this effective Hamiltonian becomes a quantum $E_{8}$ Hamiltonian,
\begin{equation}
\mathcal{H}_{E_{8}}=\mathcal{H}_{c=1/2}+h\int dx\sigma(x),
\label{S5}
\end{equation}
where $h$ corresponds to $\w_0/2$ in the scaling limit. The Schr$\ddot{\rm{o}}$dinger equation in the scaling limit becomes
\begin{equation}
i\partial_{t}\mathcal{U}_{D,s}(t)\vert\psi_{s}(t)\rangle =\mathcal{H}_{E_{8}}\mathcal{U}_{D,s}(t)\vert\psi_{s}(t)\rangle
\label{S6}
\end{equation}
with $\mathcal{U}_{D,s}(t)=\exp\left\lbrace -iht\int\sigma(x)dx\right\rbrace$ being the correspondence of $\mathcal{U}_{D}(t)$ in the scaling limit. Solving this, we can obtain
\begin{equation}
\vert\psi_{s}(t)\rangle = \mathcal{U}_{D,s}^{-1}(t)e^{-i\mathcal{H}_{E_{8}}t}\vert\psi_{s}(0)\rangle,
\label{S7}
\end{equation}
where $\wec{\psi_{s}(0)}$ is the ground state of the driven Ising chain at time $t=0$ in the scaling limit.

\section{B.~post-quench perturbation framework and the singularity regularization}
\label{app:C}
In this part, we provide the necessary formalism for solving the overlap between the initial state and complete quantum $E_8$ basis via perturbation theory, known as the post-quench framework~\cite{H_ds_gi_2019}. The detailed derivations and results are well organized and performed in \cite{H_ds_gi_2019}, and here we just quote the necessary results for our calculations. The post-quench overlap between the single-$E_8$-particle basis and the perturbed wave function is given by,
\begin{equation}
\mathcal{F}_{i}={}_{a_{i}}\langle\{0\}\wec{\psi(0)} = h\frac{F^{\sigma}_{i}}{m_{a_{i}}^{2}} + h^{2} \left(\sum_{j=1}^{8}\frac{ F_{j}^{\sigma}F_{ij}^{\sigma}(i\pi,0) }{ m_{a_{i}}^{2}m_{b_{j}}^{2}  } + \text{higher form factor contributions}\right) + O(h^3),
\label{S29}
\end{equation}
where $\wec{\{0\}}$ implies that this single particle state carries 0 momentum trivially due to the momentum constraint in the perturbed wave function. The form factor $F^{\mathcal{O}}_{a_{1},...,a_{n}}$ is given by the definition as  
\begin{equation}
    F^{\mathcal{O}}_{a_{1},...,a_{n}}=\langle 0 \vert\mathcal{O}\vert\theta_{1},...,\theta_{n}\rangle_{a_{1},...,a_{n}}.
\label{S12}
\end{equation}
The basic $E_{8}$ form factor theory and a complete form factor bootstrapping process can be found in many previous works, 
for instance, Ref.~\cite{DELFINO1995724,H_ds_gi_2019,xiao}, etc.
The post-quench overlap between the two-$E_8$-particle basis and the perturbed wave function is given by,
\begin{equation}
\mathcal{F}_{ij}=\int\frac{d\theta}{2\pi}{}_{a_{i},a_{j}}\langle\{\theta,\theta_{ij}\}\wec{\psi(0)}=\int \frac{d\theta}{2\pi} K_{ij}(\theta,\theta_{ij}),
\label{S30}
\end{equation}
where $\theta_{ij}$ is given by the momentum conservation $m_{i}\sinh\theta +m_{j}\sinh\theta_{ij}=0$. The explicit form of $K_{ij}(\theta,\theta_{ij})$ is
\begin{equation}
K_{ij}(\theta,\theta_{ij})=h\frac{F^{\sigma*}_{ij}(\theta,\theta_{ij})}{C_{ij}(\theta,\theta_{ij})}+O(h^2),
\label{S31}
\end{equation}
with
\begin{equation}
C_{ij}(\theta,\theta_{ij})=(m_{i}\cosh\theta + m_{j}\cosh\theta_{ij})m_{j}\cosh\theta_{ij}.
\label{S32}
\end{equation}

\section{C.~Calculation of the time-dependent magnetization}
\label{app:D}
In the scaling limit, the time-dependent magnetization $\mathscr{M}^{\alpha}(t)$s turn to be
\begin{equation}
    \begin{aligned}
        \mathscr{M}^{\varepsilon}(t)&=\cos(\w_{0}t)\mathscr{D}^{\varepsilon}(t)+\sin(\w_{0}t)\mathscr{D}^{y}(t),\\
        \mathscr{M}^{y}(t)&=\cos(\w_{0}t)\mathscr{D}^{y}(t)-\sin(\w_{0}t)\mathscr{D}^{\varepsilon}(t),\\
        \mathscr{M}^{\sigma}(t)&=\mathscr{D}^{\sigma}(t),
    \end{aligned}
    \label{Eq:MtimedomainS}
\end{equation}
where $\mathscr{D}^{\mathcal{O}}(t)$s are given by
\begin{equation}
    \mathscr{D}^{\mathcal{O}}(t)=\sum_{r,s=0}^{\infty}\langle\psi_{s}(0)\vert r\rangle\langle r\vert e^{i\mathcal{H}_{E_{8}}t}\mathcal{O} e^{-i\mathcal{H}_{E_{8}}t}\vert s\rangle\langle s\vert\psi_{s}(0)\rangle.
    \label{Eq:Dformalism}
\end{equation}
The $\sum_{r=0}^{\infty}\vert r\rangle\langle r\vert$ and $\sum_{s=0}^{\infty}\vert s\rangle\langle s\vert$ are two $E_8$ complete bases labeled by the total account of the particles included in the basis. 
It is proved that in the post-quench overlap, only single- and two-particle basis will make significant contributions~\cite{H_ds_gi_2019}, thus in the following calculation we only include the $E_8$ complete bases with total particle number up to 2.
Since $\varepsilon$ and $\sigma$ are the primary fields of Ising conformal field theory, the exact form factors can be determined using the form factor bootstrap~\cite{DELFINO1995724,H_ds_gi_2019,xiao}. Plugging the above post-quench overlap formulas into Eq.~(\ref{Eq:MtimedomainS}) and Eq.~(\ref{Eq:Dformalism}), the results are given by
\begin{equation}
        \mathscr{D}^{\varepsilon}(t)=\sum_{r,s=0}^{2} A^{\varepsilon}_{rs}(t),
        \mathscr{D}^{\sigma}(t)=\sum_{r,s=0}^{2} A^{\sigma}_{rs}(t),
    \label{Eq:DresultsXZ}
\end{equation}
with the detailed components,
\begin{equation}
\begin{aligned}
    A_{00}^{\varepsilon}(t)&=\langle\varepsilon\rangle, A_{00}^{\sigma}(t)=\langle\sigma\rangle, \\
    A_{01}^{\varepsilon}(t)&=\sum_{i}e^{-i m_{i}t} F^{\varepsilon}_{i}\mathcal{F}_{i}, A_{01}^{\sigma}(t)=\sum_{i}e^{-i m_{i}t} F^{\sigma}_{i}\mathcal{F}_{i},\\ A_{10}^{\varepsilon}(t)&=A_{01}^{\varepsilon*}(t),A_{10}^{\sigma}(t)=A_{01}^{\sigma*}(t),\\
    A_{11}^{\varepsilon}(t)&=\sum_{i,j}e^{i(m_i-m_j)t}F^{\varepsilon}_{ij}(i\pi,0)\mathcal{F}^{*}_{i}\mathcal{F}_{j}, A_{11}^{\sigma}(t)=\sum_{i,j}e^{i(m_i-m_j)t}F^{\sigma}_{ij}(i\pi,0)\mathcal{F}^{*}_{i}\mathcal{F}_{j},\\
    A_{02}^{\varepsilon}(t)&=\sum_{a,b}\int \frac{d\theta}{2\pi} e^{-iE_{ab}t}F^{\varepsilon}_{ab}(\theta_{a},\theta_{ab}) K^{\sigma}_{ab}(\theta_{a},\theta_{ab}),A_{02}^{\sigma}(t)=\sum_{a,b}\int \frac{d\theta}{2\pi} e^{-iE_{ab}t}F^{\sigma}_{ab}(\theta_{a},\theta_{ab}) K^{\sigma}_{ab}(\theta_{a},\theta_{ab}),\\
    A_{20}^{\varepsilon}(t)&=A_{02}^{\varepsilon*}(t),A_{20}^{\sigma}(t)=A_{02}^{\sigma*}(t),\\
    A_{12}^{\varepsilon}(t)&=\sum_{r,a,b}\int \frac{d\theta_{a}}{2\pi} e^{i(m_{r}-E_{ab})t}F^{\varepsilon}_{rab}(i\pi,\theta_{a},\theta_{ab})\mathcal{F}^{*}_{r} K^{\sigma}_{ab}(\theta_{a},\theta_{ab}),\\
    A_{12}^{\sigma}(t)&=\sum_{r,a,b}\int \frac{d\theta_{a}}{2\pi} e^{i(m_{r}-E_{ab})t}F^{\sigma}_{rab}(i\pi,\theta_{a},\theta_{ab})\mathcal{F}^{*}_{r} K^{\sigma}_{ab}(\theta_{a},\theta_{ab}),\\
    A_{21}^{\varepsilon}(t)&=A_{12}^{\varepsilon*}(t),A_{21}^{\sigma}(t)=A_{12}^{\sigma*}(t),\\
    A_{22}^{\varepsilon}(t)&=\sum_{r,s,a,b}\int \frac{d\theta_{r}}{2\pi}\int \frac{d\theta_{a}}{2\pi} e^{i(E_{rs}-E_{ab})t}F^{\varepsilon}_{rsab}(\theta_{r}+i\pi,\theta_{rs}+i\pi,\theta_{a},\theta_{ab})K^{\sigma*}_{rs}(\theta_{r},\theta_{rs}) K^{\sigma}_{ab}(\theta_{a},\theta_{ab}),\\
    A_{22}^{\sigma}(t)&=\sum_{r,s,a,b}\int \frac{d\theta_{r}}{2\pi}\int \frac{d\theta_{a}}{2\pi} e^{i(E_{rs}-E_{ab})t}F^{\sigma}_{rsab}(\theta_{r}+i\pi,\theta_{rs}+i\pi,\theta_{a},\theta_{ab})K^{\sigma*}_{rs}(\theta_{r},\theta_{rs}) K^{\sigma}_{ab}(\theta_{a},\theta_{ab}).
    \label{Eq:DresultsXZdetail}
\end{aligned}
\end{equation}
The vacuum expectation values $\langle\varepsilon\rangle$ and $\langle\sigma\rangle$ are related with the rotation frequency $\omega_0$, and can be obtained from the iTEBD simulation. $m_{i}$s are the mass of the quantum $E_8$ particles in unit of the mass of the lightest particle $m_{1}$, where $m_1$ can be determined through the scaling relation $m_1=C \omega_{0}^{8/15}$. $E_{ab(rs)}$ are the total energy of a two particle complete basis, given by $E_{ab(rs)}=m_{a(r)}\cosh\theta_{a(r)} +m_{b(s)}\cosh\theta_{ab(rs)}$, together with the momentum conservation $0=m_{a(r)}\sinh\theta_{a(r)} +m_{b(s)}\sinh\theta_{ab(rs)}$.
The only difficulty would then be the calculation of $\mathscr{D}^{y}(t)$. By considering the lattice formalism, we have,
\begin{equation}
    \frac{d D^{z}(t)}{dt}=\frac{d}{dt}\langle\psi(0)|e^{iH_{\text{eff}}t} \sigma^{z} e^{-iH_{\text{eff}}t}|\psi(0)\rangle =i \langle\psi(0)|e^{iH_{\text{eff}}t} \left[ H_{\text{eff}}, \sigma^{z} \right]e^{-iH_{\text{eff}}t}|\psi(0)\rangle=2D^{y}(t).
\end{equation}
Sending the system back to the scaling limit, we have 
\begin{equation}
    \mathscr{D}^{y}(t)=\frac{1}{2}\frac{d \mathscr{D}^{\sigma}(t)}{dt}.
    \label{Eq:AresultsY}
\end{equation}

\section{D.~Correspondence between lattice formalism and field theory}
\label{app:E}
In the quantum $E_8$ field theory as Eq.~(\ref{S5}), an exact scaling relation between the longitudinal field deformation $h$ and the mass of the lightest $E_8$ particle $m_1$ is given by $m_1 = C_{s} h^{8/15}$, obtained by V. A. Fateev in 1994~\cite{fateev} for the first time, where the exact value of $h$ is given as $h=4.40490858...$. Such scaling relation also holds for the lattice Hamiltonian as the small longitudinal field perturbed critical transverse field quantum Ising chain~[Eq.~(\ref{S4})], given by $m_{1}=C_{\w} h^{8/15}_{z}$. Here $h_{z}=\w_{0}/2$ for Eq.~(\ref{S4}), and the coefficient $C_{\w}$ differs with $C_{s}$ due to the essential lattice formalism. For solving the new coefficient $C_{\w}$, let's start from a more general lattice Hamiltonian,
\begin{equation}
    \mathcal{H}_{\text{lat}}=-J\sum_{i}\left(\sigma^{z}_{i}\sigma^{z}_{i+1}+g\sigma^{x}_{i}+h_{z}\sigma^{z}_{i}\right),
\end{equation}
with $g\lesssim 1$ and $h_{z}\rightarrow 0$. Taking the scaling limit, the corresponding field theory Hamiltonian reads
\begin{equation}
    \mathcal{H}_{s}=H_{c=1/2}+(1-g)\int \epsilon(x)dx+h\int\sigma(x)dx.
\end{equation}
For connecting the longitudinal field deformation between the lattice formalism and the field theory, we consider the correspondence between the two expectation values as
\begin{equation}
    \langle h\int\sigma(x)dx\rangle \Leftrightarrow \langle Jh_{z}\sum_{i}\sigma_{i}^{z}\rangle.
\end{equation}
For the field theory expectation value,
\begin{equation}
    \langle h\int\sigma(x)dx\rangle =h L \bar{\sigma}=hL\bar{s}M^{1/8},
    \label{Eq:fieldh}
\end{equation}
where $L=Na\rightarrow\infty$ is the length of the system, with $a$ being the lattice spacing. $\bar{\sigma}$ is the expectation value of the spin density operator $\sigma$, $M=2J(1-g)$, and $\bar{s}=2^{1/12}e^{-1/8}A^{3/2}$, $A=1.2824271291...$ known as Glaisher's constant~\cite{zam2003}. During the process for taking the scaling limit, we have chosen the convention as $2Ja=1$. For the lattice expectation value, since $\langle\sigma^{z}\rangle=(1-g^2)^{1/8}$~\cite{sachdev_2011}, we have
\begin{equation}
    \langle Jh_{z}\sum_{i}\sigma_{i}^{z}\rangle=Jh_{z}N\langle\sigma^{z}\rangle=Jh_{z}N(1-g^2)^{1/8}.
    \label{Eq:latticehz}
\end{equation}
Combining Eq.~(\ref{Eq:fieldh}) and Eq.~(\ref{Eq:latticehz}), we obtain the correspondence between $h$ and $h_{z}$,
\begin{equation}
    h=\frac{2}{\bar{s}}J^{15/8}h_{z}.
    \label{Eq:handhz}
\end{equation}
Considering the correspondence of the scaling relation, we have $m_{1}=C_{l} h^{8/15}_{z}=C_{s} h^{8/15}$, thus
\begin{equation}
    C_{l}=C_{s}J (\frac{2}{\bar{s}})^{8/15}=5.4154...~.
\end{equation}
Since $h_{z}=\w_0 /2$, for the lattice Hamiltonian Eq.~(\ref{S4}), the scaling relation is give by
\begin{equation}
    m_{1}=C_{\w}\w^{8/15}_{0}
    \label{Eq:scalingm}
\end{equation}
with $C_{\w}=C_{l}/2^{8/15}=3.7418...$ . Similarly, in the quantum $E_8$ field theory, it was determined that~\cite{DELFINO1995724}
\begin{equation}
    \langle\sigma\rangle = 1.27758... h^{1/15}. 
\end{equation}
On the lattice, we can determine
\begin{equation}
    \langle\sigma^{z}\rangle = \bar{s}^{-1}J^{-1/8} \langle\sigma\rangle = \bar{s}^{-1}J^{-1/8} \left(\frac{2}{\bar{s}}J^{15/8}\frac{\w_0}{2}\right)^{1/15} = C_{\sigma} {\w_0}^{1/15}
    \label{Eq:scalingz}
\end{equation}
with $C_{\sigma}=0.9219...$~. The scaling relation of $\langle\sigma^{x}\rangle$ with $\w_0$ can only be determined with iTEBD algorithm, as
\begin{equation}
    \langle\sigma^{x}\rangle = \langle\sigma^{x}\rangle_{\w_0 = 0}-C_{\varepsilon} \w_0^{8/15},
    \label{Eq:scalingx}
\end{equation}
where $\langle\sigma^{x}\rangle_{\w_0 = 0}=2/\pi$ can be determined analytically~\cite{Pfeuty,jd_2018tfic}, and $C_\varepsilon = 0.2342...$ is fixed by iTEBD algorithm. The results of Eq.~(\ref{Eq:scalingm}), Eq.~(\ref{Eq:scalingz}) and Eq.~(\ref{Eq:scalingx}) are used for normalization of the field theory results and comparison with lattice simulations.

\section{E.~Performing iTEBD simulation}
\label{app:F}
For verification of the analytical results obtained via field theory and post-quench perturbation approach, we use infinite time-evolving block decimation (iTEBD) algorithm to realize the lattice simulation for both the original lattice model and the Rydberg lattice model. The iTEBD algorithm can simulate the ground state wave function of an infinite large one dimensional quantum system and calculate time evolution process or correlation function of certain operators. Following the lattice formalism,
\begin{equation}
    M^{\alpha}(t)=\langle\psi(0)|e^{iH_{\text{eff}}t} \mathcal{O}_{\text{lat}}^{\alpha}(t) e^{-iH_{\text{eff}}t}|\psi(0)\rangle, \alpha=x,y,z.
    \label{Eq:MlatticeSM}
\end{equation}
The effective Hamiltonian is given by Eq.~(\ref{S4}), and the corresponding operators $\mathcal{O}_{\text{lat}}^{\alpha}$ is given as following, 
\begin{equation}
\begin{aligned}
\mathcal{O}_{\text{lat}}^{x}(t)&=\cos{\w_{0}t}\sigma^{x}+\sin{\w_{0}t}\sigma^{y},\\
\mathcal{O}_{\text{lat}}^{y}(t)&=\cos{\w_{0}t}\sigma^{y}-\sin{\w_{0}t}\sigma^{x},\\
\mathcal{O}_{\text{lat}}^{z}(t)&=\sigma^{z}.
\end{aligned}
\label{Eq:OperatorSM}
\end{equation}
For performing the lattice simulation of the time-evolution process, we first generate $\wec{\psi(0)}$, the ground state of a critical transverse field quantum Ising chain, with bond dimension $\chi = 64$, and the convergence of the singular values being lower than $10^{-11}$. Then we use the effective lattice Hamiltonian with a small longitudinal field $h=0.015$ (corresponding to the rotating frequency being $\omega_0=0.03$) to do time evolution process on this wave function. and calculate the expectation values of $\mathcal{O}_{\text{lat}}^{\alpha}$ at every time point, with a fifth order Trotter-Suzuki decomposition for total time $t=1000000$ and $dt=0.1$. Furthermore, the vacuum expectation values of $\langle\sigma^{z}\rangle$ and $\langle\sigma^{x}\rangle$ of the ground state of the effective Hamiltonian Eq.~(\ref{S4}) are also used for the normalization of $\langle\sigma\rangle$ and $\langle\varepsilon\rangle$ in the quantum $E_8$ field theory.

\section{F.~Theoretical proof of the singular peaks}
\label{app:G}
Since the periodically driven time evolution process starts at $t=0$, for better revealing the low energy collective excitations of this non-equilibrium system, and comparing the analytical results with the numerical simulation, we apply a one-sided Fourier transformation to both the analytical and numerical data. By defining $\mathscr{D}^{\mathcal{O}}(\w)=\int_{0}^{\infty} dt e^{i\w t} \mathscr{D}^{\mathcal{O}}(t)$ and recalling the relation between $\mathscr{D}^{y}(t)$ and $\mathscr{D}^{\sigma}(t)$ as Eq.~(\ref{Eq:AresultsY}), it is straightforward to obtain that $\mathscr{D}^{y}(\w)=-i\w \mathscr{D}^{\sigma}(\w)/2 $.
Then, for the analytical results Eq.~(\ref{Eq:MtimedomainS}), after the one-sided Fourier transformation, we have
\begin{equation}
    \begin{aligned}
        \mathscr{M}^{\varepsilon}(\w)&=\frac{1}{2}\left[\mathscr{D}^{\varepsilon}(\w+\w_0)+\mathscr{D}^{\varepsilon}(\w-\w_0)\right]-\frac{1}{4}\left[(\w+\w_0)\mathscr{D}^{\sigma}(\w+\w_0)-(\w-\w_0)\mathscr{D}^{\sigma}(\w-\w_0)\right],\\
        \mathscr{M}^{y}(\w)&=\frac{i}{2}\left[\mathscr{D}^{\varepsilon}(\w+\w_0)-\mathscr{D}^{\varepsilon}(\w-\w_0)\right]-\frac{i}{4}\left[(\w+\w_0)\mathscr{D}^{\sigma}(\w+\w_0)+(\w-\w_0)\mathscr{D}^{\sigma}(\w-\w_0)\right],\\
        \mathscr{M}^{\sigma}(\w)&=\mathscr{D}^{\sigma}(\w),
    \end{aligned}
    \label{Eq:Mfft}
\end{equation}
By making use of the following expression for one-sided Fourier transformation, we prove that there are two types of singular peaks inside the whole spectra, as the delta-function type and the edge-singularity type.
\begin{equation}
    \int_{0}^{\infty} dt e^{i\w t} S = S\lim_{\epsilon\rightarrow 0}\int_{0}^{\infty} dt e^{(i\w -\epsilon) t} =\lim_{\epsilon\rightarrow 0}\frac{iS}{\w+i\epsilon}=\mathcal{P}\left(i\frac{S}{\w}\right)+S\pi\delta(\w).
    \label{Eq:OSFFT}
\end{equation}
where $S$ could be some constants or functions that do not have $t$ dependence. In the following we will use $\mathscr{M}^{\sigma}(\w)=\mathscr{M}^{\sigma}(\w)$ as an example to show these two types of singular peaks. 
$\mathscr{M}^{\varepsilon}(\w)$ and $\mathscr{M}^{y}(\w)$ are given by shifting of $\mathscr{M}^{\sigma}(\w)$ with the rotating frequency $\omega_0$ according to Eq.~(\ref{Eq:Mfft}). For the detailed analysis of the singular peaks, we take $\mathcal{O}=\sigma$ as an example, and the form for $\mathcal{O}=\varepsilon$ only differs with the operator.

\subsection{a. The delta-function type singular peak}
The delta-function type singular peaks are contributed by the particle channels of $(A^{\sigma}_{00}+A^{\sigma}_{01}+A^{\sigma}_{10}+A^{\sigma}_{11})(\w)$, where $A(\w)=\int_{0}^{\infty} dt e^{i\w t} A(t)$. More explicitly, for $A^{\sigma}_{00}(\w)$,
\begin{equation}
    A^{\sigma}_{00}(\w)=\mathcal{P}\left(i\frac{\langle\sigma\rangle}{\w}\right)+\langle\sigma\rangle\pi\delta(\w),
    \label{Eq:delta00}
\end{equation}
which trivially contributes to the delta-function peak at $\w=0$. For $A^{\sigma}_{01}(\w)$ and $A^{\sigma}_{10}(\w)$,
\begin{equation}
    A^{\sigma}_{01}(\w)=\sum_{i}\mathcal{P}\left(i\frac{F^{\sigma}_{i}\mathcal{F}_{i}}{\w-m_{i}}\right)+F^{\sigma}_{i}\mathcal{F}_{i}\pi\delta(\w-m_{i}), A^{\sigma}_{10}(\w)=\sum_{i}\mathcal{P}\left(i\frac{F^{\sigma}_{i}\mathcal{F}_{i}}{\w+m_{i}}\right)+F^{\sigma}_{i}\mathcal{F}_{i}\pi\delta(\w+m_{i}).
    \label{Eq:delta01}
\end{equation}
They give rise to the eight delta-function peaks with the absolute frequencies corresponding to the mass of the eight single $E_8$ particles, both in the frequency region of $\w>0$ and $\w<0$. For $A^{\sigma}_{11}(\w)$,
\begin{equation}
    A^{\sigma}_{11}(\w)=\sum_{i,j}\mathcal{P}\left(i\frac{F^{\sigma}_{ij}(i\pi,0)\mathcal{F}^{*}_{i}\mathcal{F}_{j}}{\w+m_{i}-m_{j}}\right)+F^{\sigma}_{ij}(i\pi,0)\mathcal{F}^{*}_{i}\mathcal{F}_{j}\pi\delta(\w+m_{i}-m_{j}).
    \label{Eq:delta11}
\end{equation}
For the case $i=j$, i.e., the included single $E_8$ particles being the same in the two complete bases, it will contribute to the spectral weight for the delta-function singular peak at $\w=0$. For the case $i\neq j$, i.e., different single $E_8$ particles being included in the two complete bases, it will give rise to exotic delta-function singular peaks at $\w=m_j - m_i$. All the delta-function singular peaks are the dominant structures that can be found in both the analytical and numerical results, due to the fast decaying behavior of the form factors.

\subsection{b. The edge-singularity type peak of the Van Hove singularity}
The edge-singularity type singular peaks of Van Hove's singularity are contributed by the other particle channels as $(A^{\sigma}_{02}+A^{\sigma}_{12})(\w)$ (Here only the case of $\w>0$ are considered for simplicity). It is essentially due to the divergence of the density of states near the threshold of the spectra in the frequency domain, known as Van Hove singularity~\cite{xiao,mussardo_E7}. Taking $A^{\sigma}_{02}(\w)$ as an example, we have
\begin{equation}
    A^{\sigma}_{02}(\w)= \sum_{a,b}\mathcal{P}\left(i\int \frac{d\theta_{a}}{2\pi}\frac{F^{\sigma}_{ab}(\theta_{a},\theta_{ab}) K^{\sigma}_{ab}(\theta_{a},\theta_{ab})}{\w-E_{ab}}\right) + \int \frac{d\theta_{a}}{2} F^{\sigma}_{ab}(\theta_{a},\theta_{ab}) K^{\sigma}_{ab}(\theta_{a},\theta_{ab}) \delta(\w-E_{ab}),
    \label{Eq:delta02_in}
\end{equation}
where $E_{ab}=m_{a}\cosh\theta_{a}+m_{b}\cosh\theta_{ab}$. Integrating out the delta function, we finally obtain that
\begin{equation}
    A^{\sigma}_{02}(\w)= \sum_{a,b}\mathcal{P}\left(i\int \frac{d\theta_{a}}{2\pi}\frac{F^{\sigma}_{ab}(\theta_{a},\theta_{ab}) K^{\sigma}_{ab}(\theta_{a},\theta_{ab})}{\w-E_{ab}}\right) + \frac{1}{2} \frac{F^{\sigma}_{ab}(\theta_{a},\theta_{ab})}{m_{a}m_{b}\vert\sinh(\theta_{a}-\theta_{ab})\vert} K^{\sigma}_{ab}(\theta_{a},\theta_{ab}).
    \label{Eq:delta02}
\end{equation}
When $a\neq b$, i.e., the two $E_8$ particles included in the complete bases are different, it is proved that $1/\vert\sinh(\theta_{a}-\theta_{ab})\vert \approx 1/\sqrt{\w-(m_{a}+m_{b})}$ under the additional energy conservation of $\w=m_{a}\cosh\theta_{a}+m_{b}\cosh\theta_{ab}$~\cite{xiao}. Recalling the explicit form of Eq.~(\ref{S31}), in the denominator of $K^{\sigma}_{ab}(\theta_{a},\theta_{ab})$ we have $C_{ij}(\theta,\theta_{ij})=(m_{i}\cosh\theta + m_{j}\cosh\theta_{ij})m_{j}\cosh\theta_{ij}=\w m_{j}\cosh\theta_{ij}$, which gives the final divergence behavior of the density of states as $1/\w\sqrt{\w-(m_{a}+m_{b})}$ near the threshold of $\w=m_{a}+m_{b}$. Thus for $A^{\sigma}_{02}(\w)$, when the two particles inside the complete bases are different, they give rise to an edge-singularity type of singular peaks with the singular behavior $1/\w\sqrt{\w-(m_{a}+m_{b})}$ at the threshold $\w=m_{a}+m_{b}$.

For $A^{\sigma}_{12}(\w)$, the explicit result is given by
\begin{equation}
    A^{\sigma}_{12}(\w)= \sum_{r,a,b}\mathcal{P}\left(i\int \frac{d\theta_{a}}{2\pi}\frac{F^{\sigma}_{rab}(i\pi,\theta_{a},\theta_{ab})\mathcal{F}^{*}_{r} K^{\sigma}_{ab}(\theta_{a},\theta_{ab})}{\w+m_{r}-E_{ab}}\right) + \frac{1}{2} \frac{F^{\sigma}_{rab}(i\pi,\theta_{a},\theta_{ab})\mathcal{F}^{*}_{r}}{m_{a}m_{b}\vert\sinh(\theta_{a}-\theta_{ab})\vert} K^{\sigma}_{ab}(\theta_{a},\theta_{ab}),
    \label{Eq:delta12}
\end{equation}
where the conservation of energy is given by $\w+m_{r}=m_{a}\cosh\theta_{a}+m_{b}\cosh\theta_{ab}$. For the case when $r\neq a\neq b$, by changing the variable as $\w\rightarrow\w+m_{r}$, it is straightforward to show that if $a\neq b,~m_{a}+m_{b}>m_{r}$, the edge-singularity behavior as $1/(\w+m_{r})\sqrt{\w+m_{r}-(m_{a}+m_{b})}$ at the threshold $\w=m_{a}+m_{b}-m_{r}$.

\subsection{c. The edge-singularity type singular peak as kinematic poles in form factor}
For the case of $a\neq b$ and $r=a$ (or $b$) in $A^{\sigma}_{12}(\w)$, additional edge-singular behavior comes from the form factor $F^{\sigma}_{rab}(i\pi,\theta_{a},\theta_{ab})=F^{\sigma}_{rrb}(i\pi,\theta_{r},\theta_{rb})$. Recalling the kinematic singularities of the form factor~\cite{DELFINO1995724,H_ds_gi_2019,xiao,mussardo_E7},
\begin{equation}
    -i\lim_{\theta_{a}\rightarrow\theta_{b}}(\theta_{a}-\theta_{b})F^{\mathcal{O}}_{a,b,a_{1},...,a_{n}}(\theta_{a}+i\pi,\theta_{b},\theta_{a_{1}},...,\theta_{a_{n}})=\left(1-\prod_{n}S_{b,a_{i}}(\theta_{b}-\theta_{a_{i}})\right)F^{\mathcal{O}}_{a_{1},...,a_{n}}(\theta_{a_{1}},...,\theta_{a_{n}}),
\end{equation}
$F^{\sigma}_{rrb}(i\pi,\theta_{r},\theta_{rb})$ will show an edge-singularity behavior as $1/\theta_{r}\approx 1/\sqrt{\w-m_{b}}$ near the threshold at $\w=m_{r}+m_{b}-m_{r}=m_{b}$. Together with the edge-singularity given by the Van Hove's singularity, it seems that there will be a singular behavior of $1/\w$ appears at the threshold. However, such unphysical edge-singularity peaks would not appear, since due to the exchange behavior of the form factor,
\begin{equation}
    F^{\mathcal{O}}_{a_{1},...,a_{j},a_{j+1},...,a_{n}}(\theta_{a_{1}},...,\theta_{j},\theta_{j+1},...,\theta_{a_{n}})=S_{j,j+1}(\theta_{j}-\theta_{j+1})F^{\mathcal{O}}_{a_{1},...,a_{j+1},a_{j},...,a_{n}}(\theta_{a_{1}},...,\theta_{j+1},\theta_{j},...,\theta_{a_{n}}),
\end{equation}
we will have $F^{\sigma}_{rrb}(i\pi,\theta_{r},\theta_{rb})=S_{rb}(\theta_{r}-\theta_{rb})F^{\sigma}_{rbr}(i\pi,\theta_{rb},\theta_{r})$, where $S_{j,j+1}(\theta_{j}-\theta_{j+1})$ and $S_{rb}(\theta_{r}-\theta_{rb})$ are the S-matrices for the quantum $E_8$ field theory. 
Due to the momentum conservation that $m_{r}\sinh\theta_{r}+m_{b}\sinh\theta_{rb}=0$, in the limit of $\theta_{r}\rightarrow 0$ we also have $\theta_{rb}\rightarrow 0$. Since $S_{ij}(0)=-1$ for the S-matrices, at the threshold the pair of two edge-singularities $F^{\sigma}_{rrb}(i\pi,\theta_{r},\theta_{rb})+F^{\sigma}_{rbr}(i\pi,\theta_{rb},\theta_{r})=0$ cancel with each other, and thus will not show up in the final results.

For $A^{\sigma}_{22}(\w)$, the explicit result is given by
\begin{equation}
\begin{aligned}
    A^{\sigma}_{22}(\w) &= \sum_{r,s,a,b}\mathcal{P}\left(i\int \frac{d\theta_{r}}{2\pi}\frac{d\theta_{a}}{2\pi}\frac{F^{\sigma}_{rsab}(\theta_{r}+i\pi,\theta_{rs}+i\pi,\theta_{a},\theta_{ab})K^{\sigma*}_{rs}(\theta_{r},\theta_{rs}) K^{\sigma}_{ab}(\theta_{a},\theta_{ab})}{\w+E_{rs}-E_{ab}}\right) \\
    &+ \int\frac{d\theta_{r}}{2} \frac{F^{\sigma}_{rsab}(\theta_{r}+i\pi,\theta_{rs}+i\pi,\theta_{a},\theta_{ab})}{m_{a}m_{b}m_{s}\vert\cosh\theta_{s}\sinh(\theta_{a}-\theta_{ab})\vert} K^{\sigma*}_{rs}(\theta_{r},\theta_{rs}) K^{\sigma}_{ab}(\theta_{a},\theta_{ab}),
    \label{Eq:delta22}
\end{aligned}
\end{equation}
with the energy conservation given by $\w=m_{a}\cosh\theta_{a}+m_{b}\cosh\theta_{ab}-(m_{r}\cosh\theta_{r}+m_{s}\cosh\theta_{rs})$. It seems that the integration will cancel the Van Hove's singularity.  
However, according to the kinematic singularities, in the case of $r=a$, (or $b$, or $s=a$, or $b$), additional edge-singularity arise at the threshold of $\theta_{r}=\theta_{a}\rightarrow\w_0 = m_{b}-m_{s}$ as $1/(\theta_{r}-\theta_{a})\approx 1/\sqrt{\w+m_{s}-m_{b}}$. 
Such edge-singularities cannot be canceled with the exchange property, due to the varying range of $\theta_{r}$ and $\theta_{a}$. 
With considering the Van Hove's singularity, after integration, in the case of $r=a$ and $m_{b}\geq m_{s}$ the final edge-singularity behavior is given by a $\ln(\w-\w_0)$ at the threshold $\w_0=m_{b}-m_{s}$,
which enhances the delta-function singular peaks at the same frequency.

\end{document}